\documentclass[%
 aip,
 amsmath,amssymb,
 reprint,%
]{revtex4-1}

\usepackage{graphicx}
\usepackage{dcolumn}
\usepackage{bm}
\usepackage{gensymb}
\usepackage{accents}
\usepackage[normalem]{ulem}

\usepackage[utf8]{inputenc}
\usepackage[T1]{fontenc}
\usepackage{mathptmx}
\usepackage{etoolbox}
\usepackage[separate-uncertainty,multi-part-units = single]{siunitx} 
\usepackage{amsmath}
\usepackage{amssymb}
\usepackage{pdfpages} 
\usepackage{pgffor} 

\makeatletter
\AtBeginDocument{\let\LS@rot\@undefined}
\makeatother

\def\supplementfilename{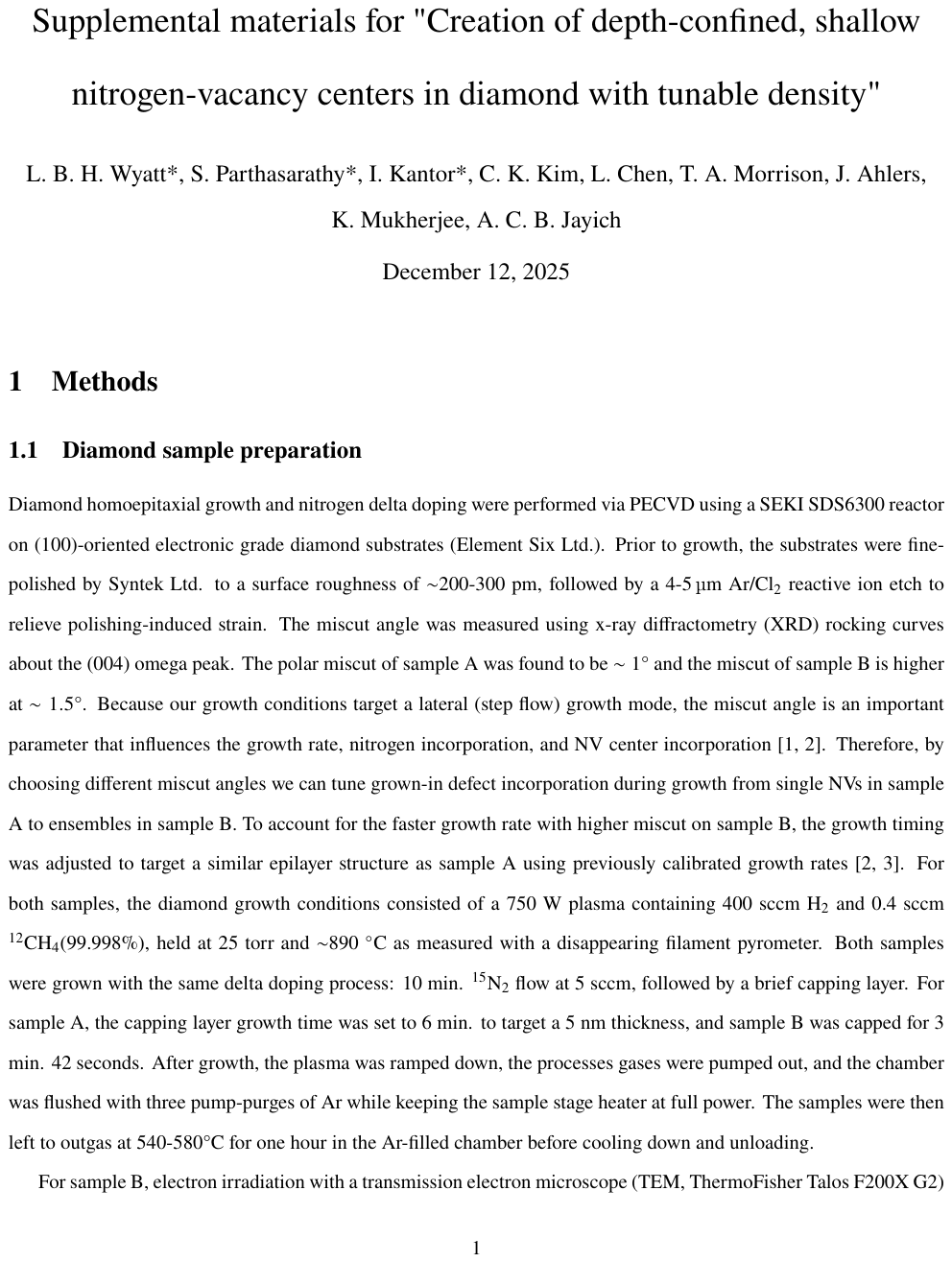}

\pdfximage{\supplementfilename}
\def\numbersupplementpages{\the\pdflastximagepages}

\newif\ifarXiv
\arXivtrue 

\usepackage{xcolor}

\makeatletter
\def\@email#1#2{%
 \endgroup
 \patchcmd{\titleblock@produce}
  {\frontmatter@RRAPformat}
  {\frontmatter@RRAPformat{\produce@RRAP{$\dagger$#1\href{mailto:#2}{#2}}}\frontmatter@RRAPformat}
  {}{}
}%
\makeatother
\def\thefootnote{*}\footnotetext{These authors contributed equally to this work.}

\begin{document}

\preprint{AIP/123-QED}

\title{Creation of depth-confined, shallow nitrogen-vacancy centers in diamond with tunable density}

\author{Lillian B. Hughes Wyatt$^*$}
\affiliation{Materials Department, University of California, Santa Barbara, Santa Barbara, CA 93106, U.S.A.}
\affiliation{Division of Engineering and Applied Science, California Institute of Technology, Pasadena, CA 91125, U.S.A.}

\author{Shreyas Parthasarathy$^*$}
\affiliation{Department of Physics, University of California, Santa Barbara, Santa Barbara, CA 93106, U.S.A.}

\author{Isaac Kantor$^*$}
\affiliation{Department of Physics, University of California, Santa Barbara, Santa Barbara, CA 93106, U.S.A.}

\author{Casey K. Kim}
\affiliation{Materials Department, University of California, Santa Barbara, Santa Barbara, CA 93106, U.S.A.}

\author{Lingjie Chen}
\affiliation{Department of Physics, University of California, Santa Barbara, Santa Barbara, CA 93106, U.S.A.}

\author{Taylor A. Morrison}
\affiliation{Department of Physics, University of California, Santa Barbara, Santa Barbara, CA 93106, U.S.A.}

\author{Jeffrey Ahlers}
\affiliation{Department of Physics, University of California, Santa Barbara, Santa Barbara, CA 93106, U.S.A.}

\author{Kunal Mukherjee}
\affiliation{Department of Materials Science and Engineering, Stanford University, Palo Alto, CA 94305, U.S.A.}

\author{Ania C. Bleszynski Jayich}
\email{ania@physics.ucsb.edu}
\affiliation{Department of Physics, University of California, Santa Barbara, Santa Barbara, CA 93106, U.S.A.}
\date{\today}

\begin{abstract}
Engineering shallow nitrogen-vacancy (NV) centers in diamond holds the key to unlocking new advances in nanoscale quantum sensing. We find that the creation of near-surface NVs through delta doping during diamond growth allows for tunable control over both NV depth confinement (with a twofold improvement relative to low-energy ion implantation) and NV density, ultimately resulting in highly-sensitive single defects and ensembles with coherence limited by NV-NV interactions. Additionally, we demonstrate the utility of our shallow delta-doped NVs by imaging magnetism in few-layer CrSBr, a two-dimensional magnet. We anticipate that the control afforded by near-surface delta doping will enable new developments in NV quantum sensing from nanoscale NMR to entanglement-enhanced metrology.
\end{abstract}

\maketitle
Solid-state quantum systems offer a promising avenue for building sensors that may be operated flexibly and robustly in a variety of device configurations. 
The nitrogen-vacancy (NV) center in diamond is a prominent example which has successfully enabled magnetic field, electric field, and temperature measurements. \cite{barry_sensitivity_2020,degen_quantum_2017} 
Near-surface NV centers placed within nanometers of a sensing target are particularly useful for sensing highly localized fields that fall off steeply with distance. 
Applications include detecting external nuclear\cite{staudacher_nuclear_2013} or electron\cite{mamin_detecting_2012,simpson_electron_2017} spins, imaging biological and condensed matter systems,\cite{Casola2018,Schirhagl2014,rizzato_quantum_2023,tetienne_quantum_2017} and probing molecular dynamics near interfaces. \cite{zheng_probing_2025,staudacher_nuclear_2013,xu_probing_2025,liu_surface_2022} 
Enabling operation of high-sensitivity, tunable density NV centers in reliably close ($\lesssim$ \qty{10}{\nano\metre}) proximity to the diamond surface would facilitate routine nanometer-scale magnetic field imaging, strong dipole coupling to spins placed on the diamond surface, and access to dipolar-driven metrologically useful entangled states. \cite{gao_signal_2025,wu_spin_2025}

Currently, most experiments that employ shallow NV centers rely on ion implantation for defect creation due to its commercial availability and convenience. 
However, this approach faces challenges for both single and ensemble-based NV sensor fabrication that stem from variability in ion-implanted NV depths and properties. 
This variation results in a low yield of usable NVs, exacerbated by decreased coherence and charge stability near the surface. \cite{Rogers2021} 
For single-NV-based sensors, these low yields necessitate a time-consuming post-selection process when targeting a particular depth.
For ensembles, which require high implantation dosages, coherence and contrast is further degraded due to lattice damage and uncontrolled creation of other defects, resulting in severely limited sensitivity. \cite{tetienne_spin_2018} 

Nitrogen delta doping during plasma-enhanced chemical vapor deposition (PECVD) diamond growth is a promising alternative for creating high-quality shallow NV centers. 
This technique bypasses the need for implantation and has proven capable of forming coherent NV centers with significantly improved depth confinement at distances of $\gtrsim$\qty{100}{\nano\metre} from the surface, with the confinement even reaching a two-dimensional (2D) limit.\cite{hughes_two-dimensional_2023, davis_probing_2023, gao_signal_2025}
Near-surface delta doping has been investigated before,\cite{Ohno2012,osterkamp_stabilizing_2015,ohashi_negatively_2013} but each of these works measured only a few NVs and none studied depth confinement or depth-correlated properties below $\sim$ \qty{10}{\nano\metre} depth. 

\begin{figure}
\centering
\includegraphics[width=0.85\linewidth]{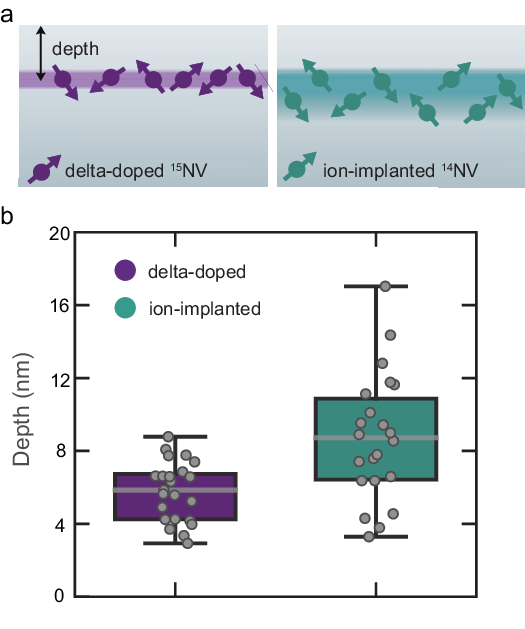}
\caption{\label{fig:depthvariance} (a) Cartoon schematic showing a delta-doped NV layer positioned as a particular depth from the diamond surface. (b) Box and whisker plot of the $^1$H NMR-measured NV depths in delta-doped (purple, sample A) and ion-implanted (teal, sample C) diamonds. The delta-doped sample A has a mean depth of \SI{5.8+-1.6}{nm} whereas the implanted sample C has a wider distribution, \SI{8.7+-3.5}{nm}.}
\end{figure}

In this letter, we demonstrate precise depth confinement and tunable NV densities $\lesssim$ \qty{10}{\nm} from the diamond surface, enabling the creation of both highly sensitive single NV centers and ensembles with NV-NV interaction-limited coherence despite their surface proximity. We use our shallow ensembles to image magnetism in few-layer CrSBr, a Van der Waals material that has attracted significant recent interest. \cite{Casola2018,ghiasi_nitrogen-vacancy_2023,tschudin_imaging_2024,melendez_quantum_2025,shao_magnetically_2025,ziebel_crsbr_2024}
We then discuss how delta-doped shallow NV samples may enable new directions in nanoscale quantum sensing. 

For this work, we prepared two diamond samples with shallow NV centers introduced via nitrogen delta doping during PECVD growth. 
Sample A was grown to contain single, individually resolvable NV centers with a \qty{5}{\nm} target depth, and sample B was grown to have a \qty{10}{\nm}-deep densely populated NV layer. 
The NV density in sample B was tuned via subsequent \qty{200}{\kilo\electronvolt} electron irradiation and annealing\cite{hughes_two-dimensional_2023} (details in SI). 
Finally, as a reference implanted sample (sample C), we use \qty{4}{\kilo\electronvolt} nitrogen ion implantation with a target depth of \qty{7}{\nm} (details and additional characterization may be found in ref.~\cite{bluvstein_identifying_2019}). 

First, we find that delta doping enables thin layer confinement of NVs near the diamond surface, as shown in Fig.~\ref{fig:depthvariance}. 
To investigate delta-doped depth confinement, we measure the depths of 23 single NVs in sample A by detecting a $^1$H nuclear magnetic resonance (NMR) signal from a layer of oil deposited atop the diamond (details on the NMR depth measurement are given in the SI).\cite{pham_nmr_2016}
We find a mean NV depth and standard deviation of \qty{5.8+-1.6}{\nm} in sample A. 
In contrast, the reference implanted sample C has a mean NV depth and standard deviation of \qty{8.7+-3.5}{\nm}, more than double the standard deviation in depth compared to sample A.
The full distribution of measured depths for both samples is depicted in Fig.~\ref{fig:depthvariance}(b).
We take sample C as a fair representation of typical shallow NV samples produced via ion implantation, and a literature survey to support this claim is presented in the SI. 

\begin{figure}
\centering
\includegraphics[width=0.85\linewidth]{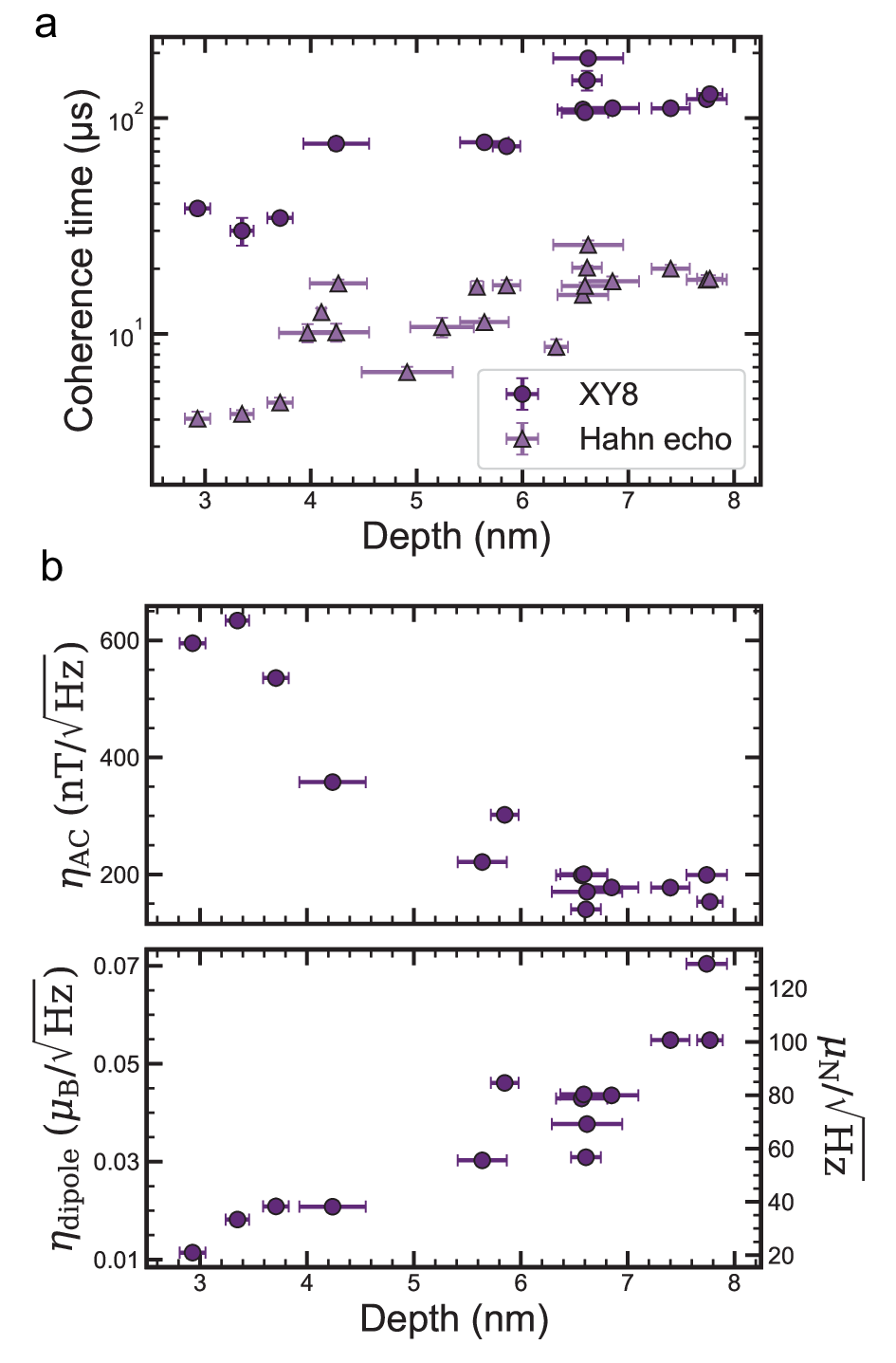}
\caption{\label{fig:coherence_sensitivity} (a) Hahn echo (light purple triangles) and saturated $T_{\textrm{2,XY8}}$ coherence (dark purple circles) versus NV depth for single NVs in delta-doped sample A. (b) Plot of $T_{\textrm{2,XY8}}$ sensitivity versus depth for sample A. Magnetic field sensitivity ($\eta_{AC}$) is shown in the top plot as calculated according to the definition derived in ref.~\cite{barry_sensitivity_2020}. Bottom plot shows sensitivities to electron and proton spins, in units of Bohr ($\mu_\textrm{B}$) and nuclear ($\mu_\textrm{N}$) magnetons.}
\end{figure}

In Fig.~\ref{fig:coherence_sensitivity}, we present the depth-registered coherence of single NV centers in sample A and use these values to calculate their AC magnetic field and dipole sensitivities.
Fig.~\ref{fig:coherence_sensitivity}(a) plots depth-dependent $T_2$, measured with Hahn echo (light purple triangles) and XY8 (dark purple circles) pulse sequences and depicts decreasing coherence with increasing surface proximity. 
Controlling for depth, the Hahn echo and XY8 coherence times presented in Fig.~\ref{fig:coherence_sensitivity}(a) are longer than most measurements of which we are aware in the literature and on par with the best implanted samples presented in ref.~\cite{sangtawesin_origins_2019}. 

Fig.~\ref{fig:coherence_sensitivity}(b) plots the calculated AC magnetic field and dipole sensitivities as a function of depth using the measured parameters of our experimental setup (see SI). 
The upper plot depicts worsening AC magnetic field sensitivity for shallower NV centers due to the onset of surface-induced decoherence. 
However, proximity to the surface also enables proximity to a sensing target. 
As a figure of merit for this benefit, we plot the calculated dipole sensitivity $\eta_\textrm{dipole}$, defined as the phase-locked sensitivity of the NV to a magnetic dipole on the diamond surface, assuming optimal coupling geometry (see SI).
Despite surface-induced decoherence, $\eta_{\textrm{dipole}}$ improves for shallower NVs. 
For the most shallow (\qty{3}{\nm} deep) NV center measured, the calculated dipole sensitivity would enable detection (with unity signal-to-noise ratio) of a single electron spin on the surface in $\sim$\qty{100}{\us}. 
We note that other commonly used spin detection methods (e.g., variance sensing of the Larmor precession of the spins) often have a stronger dependence on NV-target separation and would therefore benefit from shallower NV centers to an even greater extent.

\begin{figure}
\centering
\includegraphics[width=1.0\linewidth]{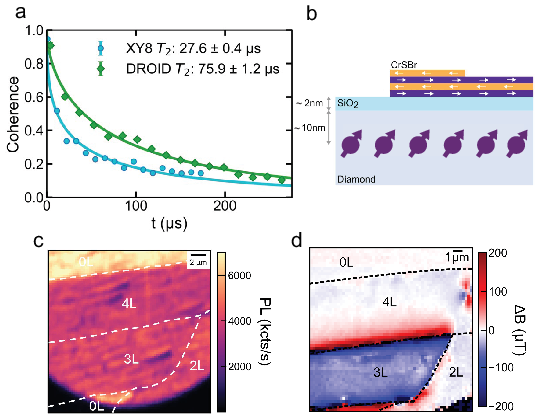}
\caption{\label{fig:CrSBr} (a) NV ensemble coherence under XY8 (light blue circles) and DROID (green diamonds) pulse sequences. The extension of $T_2$ with DROID indicates the presence of dipolar interactions in the shallow NV ensemble. (b) Cartoon schematic of a CrSBr flake above a $\sim$\qty{10}{nm} deep shallow delta-doped NV layer in diamond encapsulated with a thin layer of SiO$_2$. (c) Confocal photoluminescence image and (d) magnetic image showing stray fields arising from inter-layer ordering in CrSBr. Labels (0L, 2L, 3L, 4L) indicate the number of CrSBr layers in different regions of the flake.}
\end{figure}

We now turn to characterizing the coherence properties of high-density, shallow, delta-doped ensembles, motivated by recent progress in reaching dipolar-interaction limited coherence in the bulk.\cite{hughes_strongly_2025} 
To evaluate the limits to coherence in delta-doped shallow ensembles, we compare the effects of two types of dynamical decoupling sequences, XY8 and DROID, on sample B.
An XY8 sequence decouples low-frequency magnetic noise, and the ensemble coherence is limited by either high-frequency noise (from \textit{e.g.}, the surface) or NV-NV dipolar interactions within the ensemble. 
In contrast, while the DROID sequence similarly decouples magnetic field noise from external sources, it can further boost coherence when NV-NV dipolar interactions are strong.\cite{zhou_quantum_2020}
The measured coherence data for both decoupling sequences are shown in Fig.~\ref{fig:CrSBr}a, where we extract a coherence time from each dataset by fitting to a stretch exponential decay $C(t) = e^{-(t/T_2)^n}$.
We find the ensemble $T_{2}$ is extended from \qty{27.6+-0.4}{\us} under XY8 to \qty{75.9+-1.2}{\us} under DROID, a 2.7-fold extension in coherence. 
We attribute this extension to NV-NV dipolar interactions, the observation of which exemplifies the low-disorder lattice and surface environment achieved via delta doping. 

Additionally, we perform a proof-of-principle sensing demonstration using ensemble sample B by placing few-layer 2D CrSBr atop the diamond surface as shown in Fig.~\ref{fig:CrSBr}(b).
CrSBr is a 2D magnet with ferromagnetic intralayer coupling and antiferromagnetic interlayer coupling below $T_{\mathrm{N\acute{e}el}} = 132 \textrm{ K}$, resulting in a net zero (nonzero) total magnetization for even (odd) layers.\cite{lopez-paz_dynamic_2022}
We image the stray magnetic field at 80 K by monitoring the frequency shift in the NV resonance as a function of position beneath the flake, observing a non-zero field under the odd layers and vanishing field under the even layers, consistent with expectations (see SI). 
The size of the non-zero field is quantitatively consistent with simulations, and is larger than prior NV ensemble-based measurements that employed deeper NV ensembles.\cite{ziffer_quantum_2024}

\begin{figure}
\centering
\includegraphics[width=0.9\linewidth]{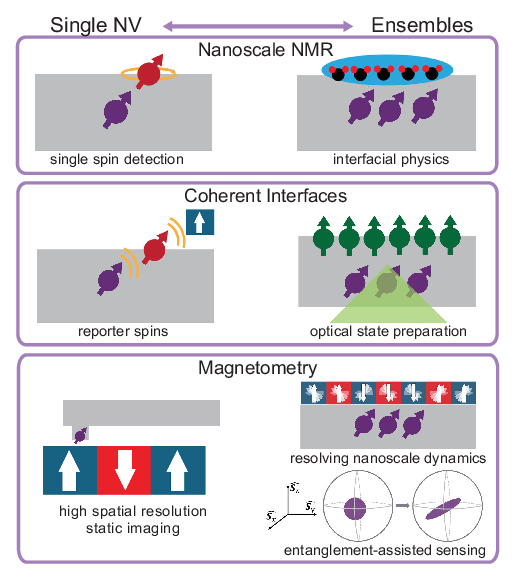}
\caption{\label{fig:outlook} Depiction of potential application spaces enabled by shallow, tunably dense NV centers. In the field of nanoscale NMR (top), single shallow NVs can couple to and detect external spins of interest,\cite{abobeih_atomic-scale_2019,biteri-uribarren_amplified_2023} while higher density ensembles may probe interfacial physics via NMR.\cite{glenn_high-resolution_2018, zheng_probing_2025} Second (middle), uniform proximity to the surface enables applications such as reporter spin-based sensing \cite{zhang_reporter-spin-assisted_2023,sushkov_magnetic_2014,schaffry_proposed_2011} and polarization transfer,\cite{rizzato_polarization_2022} where improved depth confinement consequently improves coupling homogeneity. In magnetometry (bottom), single shallow NV centers can directly image nanometer-scale features, such as magnetic domains, while depth-confined ensembles enable high-sensitivity sensing of short-wavelength fluctuations, such as those seen in antiferromagnetic phase transitions.\cite{ziffer_quantum_2024} Finally, dense two-dimensional ensembles enable entanglement-enhancements that can readily be implemented on proximal targets. }
\end{figure}

Looking ahead, in Fig.~\ref{fig:outlook} we outline a few applications that would benefit from the delta-doped shallow NV center samples we have produced here. In general, we expect both the tunability of NV density and well-controlled depth to be critical to tailoring NV systems to applications in nanoscale NMR, developing coherent quantum interfaces, and increasing sensitivity and spatial resolution in NV-based magnetometry. Mapping out nuclear spin-tagged molecules external to the diamond,\cite{abobeih_atomic-scale_2019,biteri-uribarren_amplified_2023} resolving nanoscale magnetic textures in a material,\cite{casola_probing_2018} and coherently coupling to single ``reporter" electron spins on the diamond surface\cite{zhang_reporter-spin-assisted_2023,schaffry_proposed_2011} are all challenges that involve the detection of localized sensing targets with strongly distance-dependent fields. 
In the context of NV magnetometry of condensed matter systems, shallow NVs are maximally sensitive to high-spatial-frequency features on the same order as their depth $d$. 
This high spatial resolution is useful in resolving nanometer-scale static features (with single NVs), as well as in sensing dynamical fluctuations with characteristic wavelengths on the order of $\frac{1}{d}$ (with either single spins or ensembles).\cite{casola_probing_2018}
Moreover, the NV-NV interaction-limited coherence indicates that our ensembles may saturate fundamental bounds on AC sensitivity,\cite{zhou_quantum_2020} and provide a natural test-bed for using interaction-driven entanglement to achieve even greater sensitivity or spatial resolution than is currently possible with surface or disorder-limited ensembles.\cite{gao_signal_2025,wu_spin_2025,koyluoglu_squeezing_2025}

Finally, we note that to further constrain NV positions, shallow delta doping may be readily combined with recently developed techniques for irradiation-based lateral confinement of NV centers.\cite{kim_scalable_2025} 
These approaches may facilitate higher sensitivity scanning probe devices \cite{hedrich_parabolic_2020} and lateral pre-positioning of NV centers under sensing targets. 
In summary, the shallow delta doping technique demonstrated in this work facilitates reliable engineering of highly coherent single and ensemble-based NV sensors near the surface with $\sim$\qty{1.6}{\nm} depth confinement. 

\begin{acknowledgments}
We thank Ekaterina Osipova for helpful experimental advice and troubleshooting. We also thank Samuel L. Brantly and Chenhao Jin for aid with preparing CrSBr samples.

We acknowledge support from the UC Noyce initiative, the Gordon and Betty Moore Foundation’s EPiQS Initiative via Grant GBMF10279, and the Cooperative Research on Quantum Technology (2022M3K4A1094777) through the National Research Foundation of Korea (NRF) funded by the Korean government (Ministry of Science and ICT (MSIT)). Growth studies were supported by U.S. Department of Energy BES grant No. DE-SC0019241 and sensing studies were supported by Q-NEXT, a U.S. DOE Office of Science National Quantum Information Science Research Centers under Award Number DE-FOA-0002253. 
I.K. acknowledges support from NSF QLCI program through grant number OMA-2016245 and UCSB MRL (NSF DMR-2308708).
We acknowledge the use of shared facilities of the UCSB Quantum Foundry through Q-AMASE-i program (NSF DMR-1906325), the UCSB MRSEC (NSF DMR-2308708), and the Quantum Structures Facility within the UCSB California NanoSystems Institute, and the UCSB Nanofabrication Facility, an open access laboratory.
\end{acknowledgments}

\section*{Data Availability Statement}
The data that support the findings of this study are available from the corresponding author upon reasonable request.

\section*{Author Contributions}
L. B. H. Wyatt, S. Parthasarathy, and I. Kantor contributed equally to this work.

\nocite{*}
\bibliography{ShallowNV}

\ifarXiv
   

\section*{Supplementary material}

# Supplemental materials for "Creation of depth-confined, shallow nitrogen-vacancy centers in diamond with tunable density"

L. B. H. Wyatt*, S. Parthasarathy*, I. Kantor*, C. K. Kim, L. Chen, T. A. Morrison, J. Ahlers,

K. Mukherjee, A. C. B. Jayich

December 12, 2025

## 1 Methods

### 1.1 Diamond sample preparation

Diamond homoepitaxial growth and nitrogen delta doping were performed via PECVD using a SEKI SDS6300 reactor on (100)-oriented electronic grade diamond substrates (Element Six Ltd.). Prior to growth, the substrates were fine-polished by Syntek Ltd. to a surface roughness of ~200-300 pm, followed by a 4-5 µm Ar/$Cl_2$ reactive ion etch to relieve polishing-induced strain. The miscut angle was measured using x-ray diffractometry (XRD) rocking curves about the (004) omega peak. The polar miscut of sample A was found to be $\sim 1°$ and the miscut of sample B is higher at $\sim 1.5°$. Because our growth conditions target a lateral (step flow) growth mode, the miscut angle is an important parameter that influences the growth rate, nitrogen incorporation, and NV center incorporation [1, 2]. Therefore, by choosing different miscut angles we can tune grown-in defect incorporation during growth from single NVs in sample A to ensembles in sample B. To account for the faster growth rate with higher miscut on sample B, the growth timing was adjusted to target a similar epilayer structure as sample A using previously calibrated growth rates [2, 3]. For both samples, the diamond growth conditions consisted of a 750 W plasma containing 400 sccm $H_2$ and 0.4 sccm $^{12}CH_4$(99.998%), held at 25 torr and ~890 °C as measured with a disappearing filament pyrometer. Both samples were grown with the same delta doping process: 10 min. $^{15}N_2$ flow at 5 sccm, followed by a brief capping layer. For sample A, the capping layer growth time was set to 6 min. to target a 5 nm thickness, and sample B was capped for 3 min. 42 seconds. After growth, the plasma was ramped down, the processes gases were pumped out, and the chamber was flushed with three pump-purges of Ar while keeping the sample stage heater at full power. The samples were then left to outgas at 540-580°C for one hour in the Ar-filled chamber before cooling down and unloading.

For sample B, electron irradiation with a transmission electron microscope (TEM, ThermoFisher Talos F200X G2)

was used to create locally enhanced NV concentration. The irradiation time was varied to create spots that range in dose from $10^{16}$-$10^{21}$ e$^-$/cm$^2$. Afterwards, the sample went through subsequent annealing at 400° C for 6 hours and 850° C for 6 hours in a vacuum furnace (2-4e-8 torr) to promote vacancy diffusion and NV formation. We note that electron irradiation and annealing was later applied to sample A to enhance its NV concentration, but that processing occurred after this study. Before NV characterization, the samples were cleaned in a boiling tri-acid solution (1:1:1 $H_2SO_4$:$H_2O_2$:$HClO_4$) and annealed at 450° C for 4 hours in air, followed by a piranha clean (3:1 $H_2SO_4$:$H_2O_2$, $\sim$ 140°C), rinses in water, isopropyl alcohol, and blow dried with nitrogen. Details on the preparation of sample C are given in the supplemental information of ref. [4].

## 1.2  $^1$H NMR depth measurements

NV center depths were measured according to the procedure outlined in ref. [5]. The NVs are used to detect an NMR signal produced by a statistically-polarized bath of $^1$H nuclear spins in Olympus Type F objective immersion oil deposited on the diamond surface. Fig. 1 plots the NV coherence, normalized to the background XY8 decay curve, as a function of inter-pulse spacing for two representative NVs in Sample A. We observe ambiguous resonance peaks, as shown red in in Fig. 1b, which result from the electron spin envelope modulation (ESEEM) interaction between the $^{15}$N nuclear and electronic spins and becomes exacerbated by misalignment of the bias magnetic field [6]. To suppress these ambiguous resonances, we align the bias magnetic field such that we observe >70% hyperfine polarization of the $^{15}$N nuclear spin, and we exclude any remaining ambiguous peaks from the fitting of the NMR resonance, restricting the fit to match the expected $^1$H Larmor frequency for the measured field strength.

# 2  Ion-implantation depth references

Several recent single-NV studies produced by low-energy ion implantation report depth distributions and characteristic spreads that are consistent with sample C. As noted in the main text, the mean NV depth measured in sample C (across 22 NVs, n=22) is 8.7 nm with a standard deviation of 3.5 nm. For comparison, we have estimated the standard deviations of depth distributions provided in the works of ref. [7] (n = 31) and ref. [8] (n = 21), which used nearly identical post-implantation processing as sample C, to be $\sim$5.6 nm and 4.5 nm, respectively. We also estimate the standard deviation for the reference sample in Fig. 2 of ref. [9] to be $\sim$3.5 nm. Therefore, we conclude that the scatter of depths in our ion-implanted sample C is either representative or (in the worst case) an underestimate of the scatter in depths formed via low-energy ion implantation and substrate preparation procedures currently in use in the literature.

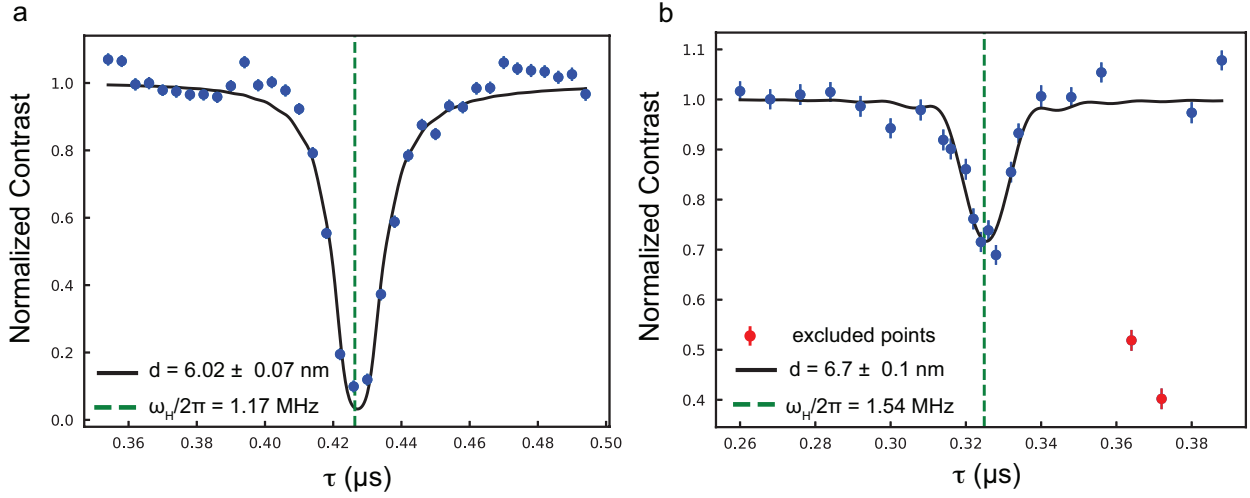

Supplementary Figure 1: $^1$H NMR depth measurements taken on two different single NVs in sample A using an XY8-12 (a) and XY8-6 (b) sequence, where 12 and 6 indicate the number of repetitions of the 8-pulse XY8 subsequence. Contrast relative to the background $T_{2,XY8}$ decay is plotted as a function of interpulse spacing $\tau$. The interpulse spacing where an NMR dip is expected is $\tau_H = \frac{\pi}{\omega_H}$, where $\omega_H$ is the $^1$H Larmor frequency calculated from the measured magnetic field, and is indicated by a green dashed line in each plot. Depths are extracted according to the model in ref. [5]. In (b) the red points are due to a $^{15}$N ESEEM resonance, and are consequently excluded from the depth-extraction fitting procedure [6].

## 3   Sensitivity Calculations

To estimate the AC magnetic field sensitivity of a single shallow NV center, we follow the shot noise limited, dynamically decoupled sensitivity expression in ref. [10].

$$\eta_{AC} \approx \frac{\pi \hbar}{2\Delta m_s g_e \mu_B} \frac{1}{\sqrt{N\tau}} e^{(\tau/T_2)^p} \sqrt{1 + \frac{1}{C^2 n_{avg}}} \sqrt{\frac{T_{overhead} + \tau}{\tau}}, \tag{1}$$

where $N = 1$ for the single-NV measurements reported here. The quantities in Eq. (1) are defined as follows: $\Delta m_s$ is the spin-projection quantum-number difference, $g_e$ is the electron $g$-factor, $\mu_B$ is the Bohr magneton, $C$ is the optical readout contrast prior to phase accumulation, $n_{avg}$ is the average number of detected photons per readout, $T_{overhead}$ is the overhead time associated with the measurement (duration of spin state initialization and readout via laser and finite duration of dynamical decoupling $\pi$ pulses), $T_2$ is the coherence time, $\tau$ is the optimal total free precession time of the measurement, and $p$ is the stretched-exponential decay exponent, taken here to be $2/3$. The final factor accounts for experimental overhead, and the expression assumes phase-locking between the dynamical-decoupled NV center and the AC magnetic field.

In our measurements, no explicit effort was made to optimize either the optical collection efficiency or the overhead times. As a result, the AC sensitivities reported in the main text represent conservative estimates of the performance achievable from these shallow, delta-doped NV centers. With improved photon collection and reduced overhead, these

sensitivities may be significantly improved.

To calculate the dipole sensitivity presented in the main text, we evaluate the field at the NV produced via dipolar interaction with a single spin located on the diamond surface, a distance $d_{\mathrm{NV}}$ above the NV:

$$B_{\mathrm{dip}} = \frac{\mu_0}{4\pi} \frac{\mu_{\mathrm{dip}}}{d_{\mathrm{NV}}^3} \left(2\cos\theta_{\mathrm{NV}}\cos\theta_{\mathrm{dip}} - \sin\theta_{\mathrm{NV}}\sin\theta_{\mathrm{dip}}\cos\phi\right), \tag{2}$$

where $\theta_{\mathrm{NV}}$, $\theta_{\mathrm{dip}}$ are the angles that the NV spin and surface dipole spin make with respect to the axis connecting them respectively, and $\phi$ is the dihedral angle between the two planes defined by the spin orientations and the axis connecting them. $\mu_{\mathrm{dip}}$ is the magnetic moment of the surface spin (either a Bohr or nuclear magneton). Optimization of the dipole orientation ($\theta_{\mathrm{dip}}$ and $\phi$) to maximize $B_{\mathrm{dip}}$ yields the maximal dipolar coupling for an NV of orientation $\theta_{\mathrm{NV}}$ relative to the quantization axis of the dipole being detected:

$$B_{\mathrm{dip}}^{\max}(\theta_{\mathrm{NV}}) = \frac{\mu_0}{4\pi} \frac{\mu_{\mathrm{dip}}}{d_{\mathrm{NV}}^3} \left(\sqrt{3\cos^2\theta_{\mathrm{NV}} + 1}\right), \tag{3}$$

For a spin placed directly above the NV on a (100) surface in a moderate (relative to the NV zero field splitting) bias field, $\theta_{\mathrm{NV}}$ is fixed to 54.7° due to the B-field independent NV quantization axis. In such a scenario, one finds that the sensitivity of an NV to the local field produced by a dipole of magnetic moment $\mu_{\mathrm{dipole}}$ can be expressed as a modification of the standard AC magnetic field sensitivity:

$$\eta_{\mathrm{dipole}} = \eta_{\mathrm{AC}}\, F_{\mathrm{dip}}(d_{\mathrm{NV}}, \mu_{\mathrm{dip}}), \tag{4}$$

where

$$F_{\mathrm{dip}}(d_{\mathrm{NV}}, \mu_{\mathrm{dip}}) = \frac{4\pi d_{\mathrm{NV}}^3}{\mu_0 \mu_{\mathrm{dip}} \sqrt{3\cos^2\theta_{\mathrm{NV}} + 1}} = \frac{4\pi d_{\mathrm{NV}}^3}{\mu_0 \mu_{\mathrm{dip}} \sqrt{2}}. \tag{5}$$

This conversion assumed an optimally oriented dipole coupling, the ability to do phase-locked coherent detection of the dipolar field (e.g., via driving the dipole concurrently), and a dipole correlation time much longer than the sensing time. Under these idealized conditions, $\eta_{\mathrm{dipole}}$ provides a convenient figure of merit for comparing the spin-detection performance of shallow NV centers.

# 4    CrSBr Sample

## 4.1    CrSBr Sample Preparation

Before transfer, diamond sample B was encapsulated with 2 nm of SiO$_2$ via atomic layer deposition to prevent possible interactions between CrSBr and the diamond surface. The CrSBr sample was then transferred onto the diamond via

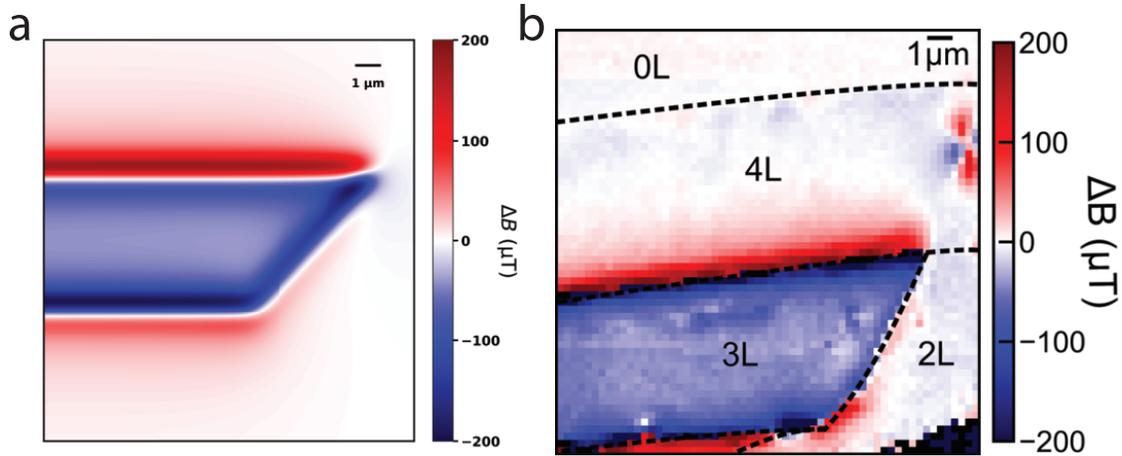

Supplementary Figure 2: (a) Simulated magnetic field projection along the NV axis using CrSBr material constants, accounting for diffraction-limited resolution of the stray field. (b) Plot from Fig. 3 in main text, reproduced for comparison to simulations

standard Van Der Waals dry-transfer protocols using a polybisphenol carbonate film on a polydimethylsiloxane support within an argon atmosphere. CrSBr layer count was identified via optical reflection contrast (previously calibrated via atomic force microscopy, Asylum Research, Cypher S AFM).

## 4.2 CrSBr Field Analysis

To understand if the stray field detected by the shallow NV centers in Fig. 3 of the main text is commensurate with expectations for 10 nm deep NV centers, we simulate the field produced by CrSBr using `magpylib` [11]. Specifically, we simulate the projection of the B field along the NV axis produced by an in-plane magnetized slab (where the magnetization direction is parallel to the projection of the NV axis onto the slab plane), at a vertical offset of 12 nm (10 nm deep NVs + 2 nm cap, as indicated in the main text). The input magnetization is calculated using the magnetic moment density expected for CrSBr from literature values of the lattice constants and magnetic moments per formula unit [12]. The field as a function of lateral position is then convolved with a gaussian function to simulate the effect of diffraction limited optics (0.7 NA objective with 532 nm wavelength). The results are in quantitative agreement with experiment, as can be seen in Fig. 2. Fig. 2(a) plots the simulated magnetic field using the procedure outlined above, and Fig. 2(b) reproduces the measured data from the main text for comparison.


\fi

\end{document}
\end{document}